\input harvmac
\input epsf
\noblackbox
\def\npb#1#2#3{{\it Nucl.\ Phys.} {\bf B#1} (19#2) #3}
\def\plb#1#2#3{{\it Phys.\ Lett.} {\bf B#1} (19#2) #3}
\def\prl#1#2#3{{\it Phys.\ Rev.\ Lett.} {\bf #1} (19#2) #3}

\def\atmp#1#2#3{{\it Adv.\ Theor.\ Math.\ Phys.} {\bf #1} (19#2) #3}
\def\jhep#1#2#3{{\it JHEP\/} {\bf #1} (19#2) #3}
\newcount\figno
\figno=0
\def\fig#1#2#3{
\par\begingroup\parindent=0pt\leftskip=1cm\rightskip=1cm\parindent=0pt
\baselineskip=11pt
\global\advance\figno by 1
\midinsert
\epsfxsize=#3
\centerline{\epsfbox{#2}}
\vskip 12pt
{\bf Fig.\ \the\figno: } #1\par
\endinsert\endgroup\par
}
\def\figlabel#1{\xdef#1{\the\figno}}
\def\encadremath#1{\vbox{\hrule\hbox{\vrule\kern8pt\vbox{\kern8pt
\hbox{$\displaystyle #1$}\kern8pt}
\kern8pt\vrule}\hrule}}

\def\frac#1#2{{#1 \over #2}}

\def\p{\partial}
\def\semi{\subset\kern-1em\times\;}

                   \def\CL{{\cal L}}

%

%

\Title{\vbox{\baselineskip12pt
\hbox{hep-th/0002117}
\hbox{EFI-2000-4}
\vskip-.5in}}
{\vbox{\centerline{D-Branes as Unstable Lumps} 
\bigskip
\centerline{in Bosonic Open String Field Theory}}}
\medskip\bigskip
\centerline{Jeffrey A. Harvey  and Per Kraus }
\bigskip\medskip
\centerline{\it Enrico Fermi Institute and Department of Physics}
\centerline{\it  University of Chicago, 
Chicago, IL 60637, USA}
\centerline{\tt harvey, pkraus@theory.uchicago.edu}
\medskip
\baselineskip18pt
\medskip\bigskip\medskip\bigskip\medskip
\baselineskip16pt
We construct Dp-branes in bosonic string theory as unstable lumps in a truncated
string field theory of open strings on a D25-brane. We find that the
lowest level truncation gives good
quantitative agreement with the predicted D-brane tension and low-lying
spectrum of the D-brane for sufficiently large $p$ and study the effect
of the next level corrections for $p=24$.  We 
show  that a $U(1)$ gauge field zero mode on the D-brane
arises through a mechanism reminiscent of the Randall-Sundrum
mechanism for gravity. 
\Date{February, 2000}
\lref\sred{M.~Srednicki,
``IIB or not IIB,''
JHEP {\bf 9808}, 005 (1998),
hep-th/9807138.}
\lref\yi{P.~Yi,
``Membranes from five-branes and fundamental strings from Dp branes,''
Nucl.\ Phys.\  {\bf B550}, 214 (1999),
hep-th/9901159.}
\lref\ks{V .A. Kostelecky and S.Samuel, ``On a Nonperturbative Vacuum for the
Open Bosonic String,'' Nucl.Phys. {\bf B336} (1990) 263.}
\lref\sz{A. Sen and B. Zwiebach, ``Tachyon Condensation in String
Field Theory,'' hep-th/9912249.}
\lref\sendesc{A. Sen, ``Descent Relations Among Bosonic D-branes,'' Int.J. Mod.
Phys. {\bf A14} (1999) 4061, hep-th/9902105.}
\lref\sena{A.~Sen, ``Stable non-BPS bound states of BPS D-branes,''
\jhep{9808}{98}{010}, hep-th/9805019; 
``SO(32) spinors of type I and other solitons on brane-antibrane pair,''
\jhep{9809}{98}{023}, hep-th/9808141;
``Type I D-particle and its interactions,''
\jhep{9810}{98}{021}, hep-th/9809111; 
``Non-BPS states and branes in string theory,'' 
hep-th/9904207, and references therein.}
\lref\sennon{A. Sen, ``BPS D-branes on non-supersymmetric cycles,'' 
\jhep{9812}{98}{021}, hep-th/9812031.}
\lref\bergman{O.~Bergman and M.~R.~Gaberdiel,
``Stable non-BPS D-particles,'' \plb{441}{98}{133}, hep-th/9806155.}
\lref\hhk{J. A. Harvey, P.Ho\v rava and P. Kraus, 
``D-Sphalerons and the Topology
of String Configuration Space,'' hep-th/0001143.}
\lref\berk{N. Berkovits, ``The Tachyon Potential in Open Neveu-Schwarz String
Field Theory,'' hep-th/0001084.}
\lref\polchinski{J.~Polchinski, ``Dirichlet-Branes and Ramond-Ramond 
Charges,'' \prl{75}{95}{4724}, hep-th/9510017.}
\lref\senspinors{A.~Sen, ``$SO(32)$ Spinors of Type I and Other Solitons on 
Brane-Antibrane Pair,'' \jhep{9809}{98}{023}, hep-th/9808141.}
\lref\phk{P. Ho\v rava, ``Type IIA D-Branes, K-Theory, and Matrix Theory,'' 
\atmp{2}{99}{1373}, hep-th/9812135.}
\lref\yi{P. Yi, ``Membranes from Five-Branes and Fundamental Strings from
D$p$-Branes,'' \npb{550}{99}{214}; hep-th/9901159.}
\lref\senpuz{A. Sen, ``Supersymmetric World-volume Action for Non-BPS
D-branes,'' \hfill\break hep-th/9909062.}
\lref\taylor{W. Taylor, ``D-brane effective field theory from string
field theory,'' hep-th/0001201.}
\lref\cft{C. G. Callan, I. R. Klebanov, A. W. Ludwig and J.M Maldacena,
``Exact solution of a boundary conformal field theory,'' Nucl.Phys.
{\bf B422} (1994) 417, hep-th/9402113;
J. Polchinski and L. Thorlacius, ``Free fermion representation of a boundary
conformal field theory,''Phys. Rev. {\bf D50} (1994) 622, hep-th/9404008;
P.Fendley, H. Saleur and N. P. Warner, ``Exact solution of a massless
scalar field with a relevant boundary interaction,'' Nucl. Phys.
{\bf B430} (1994) 577, hep-th/9406125;
A. Recknagel and V. Schomerus, ``Boundary deformation theory and
moduli spaces of D-branes,'' Nucl. Phys. {\bf B545} (1999) 233, 
hep-th/9811237.}
\lref\rs{L.Randall and R. Sundrum, ``An alternative to compactification,''
Phys.Rev.Lett. {\bf 83} (1999) 4690, hep-th/9906064.}
\lref\yi{P.Yi, ``Membranes from five-branes and fundamental strings
from Dp branes,'' Nucl.Phys.{\bf B550} (1999) 214, hep-th/9901159.}
\lref\zrefs{C.G.Callan, J.A. Harvey and A.Strominger, ``Worldbrane
actions for string solitons,'' Nucl. Phys. {\bf B367} (1991) 60;
T. Awadi, M. Cederwall, U. Gran, B.E.W. Nilsson and B. Razaznejad,
``Goldstone Tensor Modes,'' JHEP 9902:001 (1999), hep-th/9811145.}
\lref\witt{E. Witten, ``Noncommutative Geometry and String Field
Theory,'' Nucl. Phys. {\bf B268} (1986) 253.}

\newsec{Introduction}
In a remarkable paper, Kostelecky and Samuel studied tachyon condensation in
Witten's version of bosonic open string field theory \witt\ 
using a level truncation scheme and found
what appears to be rapid convergence as higher levels are included \ks.
This calculation was recently reinterpreted and extended in \sz. Following
previous arguments \sena, Sen and Zwiebach argued that this calculation 
describes
the decay of a space filling $D25$-brane to the bosonic string vacuum and found
that the energy of tachyon condensation cancels the $D25$-brane tension to
an accuracy of $99 \%$ when terms in the tachyon potential up to level 8
are included. The level truncation scheme has been further extended
in \taylor\ and tachyon condensation has been studied to lowest order
in open superstring field theory in \berk. 

Based on studies of tachyon condensation in conformal field theory \cft,
one also expects to be able to describe $Dp$-branes for $p<25$ as unstable
configurations of the open string tachyon on a space-filling $D25$-brane
\sendesc. It has been suggested that the level truncation scheme
might be a useful tool for studying this question \sz. Such an analysis
probes much more of the structure of open string field theory than
a study of the tachyon vacuum energy since it involves the higher
derivative terms present in string field theory in a non-trivial way.

In this paper we begin an investigation of $Dp$-branes in level truncated
open string field theory. We find that for $p=24$ the lowest level terms
give a remarkably good description of both the qualitative and quantitative
structure of the $D24$-brane. For smaller $p$, higher derivative terms and
higher level fields become progressively more important and  require a more
detailed analysis than will be presented here.
\newsec{$Dp$-branes at level $0$}

We will follow the conventions of \ks.  For convenience we set $\alpha'=1$, 
$g=2$, as in \sz\ with $g$ the open string coupling constant.  
In these conventions a $Dp$-brane has tension
\eqn\tension{T_p=\half (2 \pi)^{23-p}=(2\pi)^{25-p}~T_{25}.}
%

The spectrum of open strings on the $D$-brane includes a tachyon with 
$m^2 = -1$, a massless $U(1)$ gauge field, massless scalars corresponding to 
translational zero modes, and additional massive states.  

Kostelecky and Samuel used open string field theory to work out the 
$D25$-brane action to the first few levels.  Truncating at level 0 leaves
just the tachyon $\phi$ with action
\eqn\tachact{S= 2\pi^2 T_{25} \int \! d^{26}x \, \left(
{1 \over 2} \partial_\mu \phi \partial^\mu \phi - {1 \over 2}\phi^2
+ 2 \kappa \tilde{\phi}^3 \right),}
where
\eqn\defs{\kappa = {1 \over 3!}\left({3 \sqrt{3} \over 4}\right)^3, \quad
\tilde{\phi} = \exp{\left[ \ln{(3\sqrt{3}/4)}\partial_\mu  \partial^\mu\right]}
\phi.}
We are using a metric with signature 
$\eta_{\mu\nu} = {\rm diag}(-,+,+,\ldots ,
+)$.   
In the remainder of this section we will set $\tilde{\phi}=\phi$, and 
defer discussion of 
the validity of this substitution to the next section. 

The tachyon potential $V(\phi)= -{1 \over 2} \phi^2 + 2 \kappa \phi^3$
has an unstable extremum at $\phi=0$ and a locally stable extremum at
$\phi_c = 1/(6 \kappa) \approx .456$.  
The local minimum $\phi_c$ is unstable due
to nucleation of bubbles in which the  tachyon is sufficiently negative,
$\phi<\phi_*=-\phi_c/2$, where $V(\phi_*)=V(\phi_c)$.
  The nucleation of a bubble is described by a ``bounce'' --- 
a solution of the Euclidean field equations which asymptotes to $\phi_c$.  Such
a bounce solution has a single negative mode in its fluctuation spectrum.
Since the action scales like $1/g^2$, there is a natural candidate for the 
bounce: the bosonic $D$-instanton. 
This situation is to be contrasted with that in IIA string theory, which
also contains a $D$-instanton with a single negative mode.  The IIA 
D-instanton does not mediate vacuum decay, but instead signals the existence
of a non-contractible loop in the space of Euclidean IIA histories~\hhk.

Besides the D-instanton, the lowest order tachyon action supports various
$p+1$ dimensional unstable ``lump'' solutions, and it has been suggested
that  these should be identified with bosonic $Dp$-branes \sendesc.  
Consider the field equations
for static configurations with $p+1$ dimensional translation symmetry. 
Let $x^m$, $m=0 \ldots p$, be coordinates parallel to the lump, and let
$\rho$ be the radial coordinate in the $25-p$ dimensional transverse space.
Then the equation of motion for  spherically symmetric 
tachyon configurations is given by
\eqn\phieq{\partial_\rho^2 \phi + {24-p \over \rho} \partial_\rho \phi
-V'(\phi)=0.}
If we think of $\rho$ as time, then  \phieq\ can be thought of
as the equation of motion for a particle
in a potential $-V$ subject to a time dependent damping term (which
vanishes for $p=24$). 

A lump has boundary conditions
\eqn\lumpbc{\partial_\rho \phi |_{\rho=0} =0, \quad 
\lim_{\rho \rightarrow \infty} \phi(\rho) = 
\phi_c.}
For $p<24$, $\rho$ ranges from $0$ to $\infty$ and smoothness at the origin
requires that the $\rho$ derivative of $\phi$ vanish there.  For $p=24$, $\rho$
ranges from $-\infty$ to $\infty$ and the vanishing derivative at the origin
follows from the $\rho \rightarrow -\rho$ symmetry.  

For $p=24$, translation symmetry implies the conservation law
\eqn\cons{{1 \over 2}(\partial_\rho \phi)^2 - V(\phi) = -V(\phi_c) = 
{1 \over 6^3 \kappa^2},}
with solution given by
\eqn\rhosol{\rho = \int_{\phi_*}^\phi \! {d\phi' \over \sqrt{2V(\phi')-2V(\phi_c)}}.}
It does not seem possible to express $\phi(\rho)$ in closed form.   Numerical
evaluation yields the form shown by the solid line in figure 1.
 
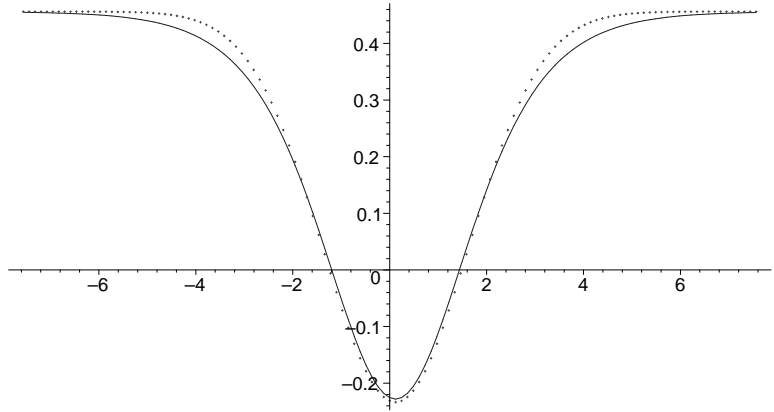
\fig{The lump solution, plotted as $\phi$ versus $\rho$.
  The solid line is the numerical solution, and the
dashed line represents the approximate fit $\phi_\ell$.}{compare.ps}{4truein}
The exact lump solution
is well approximated by the function
\eqn\applump{\phi_\ell(\rho) = \phi_c - .69 e^{-.22 \rho^2},}
as illustrated by the dashed line in figure 1.  
We will often find it convenient to work
with $\phi_\ell(\rho)$, which suffices for the accuracy needed in this paper.

The tension of the lump is given by
\eqn\lumpten{\eqalign{T^\ell_{24} =& 2 \pi^2 T_{25} \int_{- \infty}^{\infty} \! 
d\rho \,
\left\{ {1 \over 2}(\partial_\rho \phi)^2 + V(\phi)  -V(\phi_c) \right\} \cr
=& 4 \pi^2 T_{25} \int_{\phi_*}^{\phi_c} \! d\phi' \, 
\sqrt{2V(\phi')-2V(\phi_c)} \approx 4.93 T_{25} \approx .78T_{24}.\cr}}

We thus find that $T^\ell_{24}$ is $78\%$ of $T_{24}$, which  supports
the conjecture that the lump represents a $D24$-brane.

Now we turn to the description of $Dp$-branes with $p<24$.  Here we have to
proceed  numerically.  We use the shooting method, which corresponds to
tuning $\phi(0)$ such that the solution asymptotes smoothly to $\phi_c$ at
infinity.  We find the following results for $\phi(0)$ and the tension 
$T^\ell_{p}$ expressed in terms of the $Dp$-brane tension $T_p$

\eqn\eetabone{\vbox{\offinterlineskip \hrule
\halign{&\vrule#&\strut\ \  \hfil#\ \ \cr
height2pt&\omit&&\omit&&\omit&&\omit&&\omit&&\omit&&\omit&&\omit&\cr
&p~~~ &&24&&23&&22&&21&&20&&19&&$<19$&\cr
\noalign{\hrule}
height2pt&\omit&&\omit&&\omit&&\omit&&\omit&&\omit&&\omit&&\omit&\cr
&$\phi(0)$\ \ &&-.23&&-.64&&-1.5&&-3.5&&-11.5&&-$10^7$&&no sol.&\cr 
\noalign{\hrule}
height2pt&\omit&&\omit&&\omit&&\omit&&\omit&&\omit&&\omit&&\omit&\cr 
& $T^\ell_{p}/T_p$ &&.78&&.81&&.72&&.59&&.30&&.003&& no sol.&\cr}\hrule}}

The accuracy rapidly decreases with decreasing $p$.  For $p<19$ we find
no solutions; the damping term in the equation of motion prevents the 
field from making it over the hump no matter how high up the inverted
potential we take the field at the origin.  As the magnitude of $\phi(0)$
increases, the solutions develop a region of increasing $\partial_\rho \phi$
near the origin.  Thus the approximation of neglecting higher 
derivatives and higher
level fields becomes worse and worse, which accounts for the decreasing 
accuracy of the computed tension.  

\newsec{Corrections to the $D24$-brane tension}
Surprisingly, truncation to the lowest level fields and lowest number
of derivatives gives good
agreement for the tension of $Dp$-branes for sufficiently large $p$. 
There is no  {\it 
a priori} reason why this should be the case,
nor is it clear that including higher
levels and higher derivatives will yield small corrections to the lowest order 
result.   However, \refs{ \ks,\sz} found that this was the case for 
the value of the minimum of the tachyon potential (although in this case
higher derivatives played no role) and so we might hope that the same is
true here. We will make one check of this assumption by 
computing the leading higher level and derivative corrections to the tension
of the $D24$-brane.  

\subsec{Derivative correction to the tension} 

The derivative corrections arise from the appearance of $\tilde{\phi}$ in the
interaction terms.  The derivatives in the exponential are multiplied by
$\alpha'$ (which we have set to 1), and so if an $\alpha'$ expansion is 
valid then it is sensible to expand the exponential to linear order and to
see a small correction to the tension. At this order, 
this does not in fact introduce
higher derivatives into the action, but does introduce $\kappa$ dependence,
which one hopes to be an effective expansion parameter.  The action is now
\eqn\hdact{S=2\pi^2 T_{25} \int \! d^{26}x \, \left(
{1 \over 2}[1-8 \kappa\ln{(6\kappa)}\,\phi] \partial_\mu \phi \partial^\mu \phi - {1 \over 2}\phi^2
+ 2 \kappa {\phi}^3 \right).}
When $\phi$ condenses to the local minimum $\phi_c$, the coefficient of the
kinetic term, which we call $f(\phi)/2$, becomes
\eqn\kin{f(\phi_c)=1-8 \kappa\ln{(6\kappa)}\,\phi_c \approx -.05.}
The approximate vanishing of the kinetic term at the minimum seems
consistent 
with the conjecture of \senpuz~  that tachyon condensation sets all kinetic 
terms to zero and is related to the observation in 
\ks\ that there are no physical poles for the tachyon and transverse
component of the $U(1)$ gauge field in the presence of the
tachyon condensate.  After including higher 
order corrections, we expect that the vanishing of $f(\phi)$ exactly
corresponds with the minimum of the tachyon potential.

We can now solve for the corrected lump solution using
\eqn\lumpcor{{1 \over 2}f(\phi)(\partial_\rho \phi)^2 - V(\phi)=-V(\phi_c),}
which yields
\eqn\newlump{\rho = \int_{\phi_*}^{\phi} \! d\phi' \, {\sqrt{f(\phi')} \over
\sqrt{2V(\phi')-2V(\phi_c)}}.}
It only makes sense to take the upper limit of integration in the region
for which $f(\phi)$ is positive. The boundary of this region is 
$\phi_0 \approx .436$,
which is slightly less than $\phi_c$.  
 One easily finds that this translates into
a finite range for $\rho$.  Numerically, we find that the vanishing of 
$f(\phi)$ corresponds to $\rho \approx 3.15$.  The corrected lump solution
is similar to our previous results for sufficiently small $\rho$, but
differs asymptotically. The previous solution only approaches 
$\phi_c$ asymptotically, whereas in the present case the tachyon reaches
its vacuum value at finite distance.  The two solutions in the small $\rho$
region are displayed in figure 2.

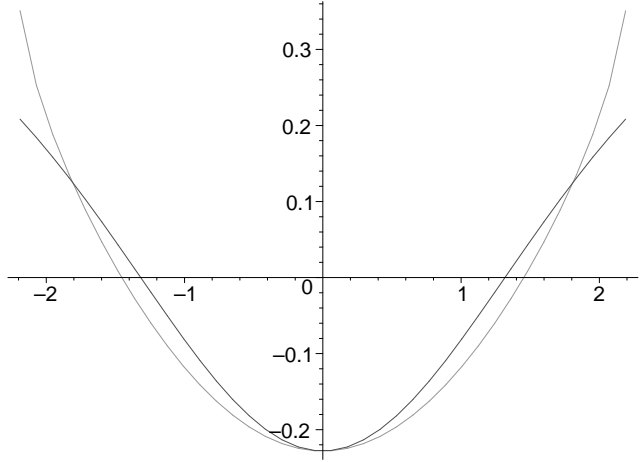
\fig{Comparison of solutions with and without the leading derivative
correction.  The corrected solution is the one developing a large 
derivative, and will reach $\phi_0$ at 
finite $\rho$.}{comparehd.ps}{3.3truein}

The computation of the $D24$-brane tension now proceeds as before, and we find
\eqn\hdten{T_{24}^\ell=4 \pi^2 T_{25} \int_{\phi_*}^{\phi_0} \! d\phi' \, 
\sqrt{1-8 \kappa\ln{(6\kappa)}\,\phi~}\sqrt{2V(\phi')-2V(\phi_c)}
\approx 4.41 \, T_{25}.}
Including the new term has decreased the tension from $78 \%$ to $70 \%$ of the
true value.   It is encouraging that the result is a small correction, albeit
in the wrong direction.   

It would be interesting to examine the results of 
including more derivative terms in the expansion of the exponential, 
but that will not
be considered here.

\subsec{Higher level corrections}

The tachyon background acts as a source for higher level fields, and so
we should examine their effects on the tension.  The action is invariant
under sign reversal of odd-level fields, which therefore must appear at
least quadratically in the action.  Thus it is consistent with the equations
of motion to set all odd-level fields to zero \sz.  We will study the effect
of the following level two fields which are excited by the tachyon 
background: an auxiliary scalar $\beta_1$, a vector $B_\mu$, and a 
symmetric tensor $B_{\mu\nu}$.  Including kinetic terms for these fields
means that we work to at least level four, and so we should include 
at least interaction terms quadratic in these fields.  In the notation
of \ks~ (see appendix B), we therefore keep $\CL^{(2)} + \CL^{(4)}$.  
The interaction terms
are an infinite series in derivatives.  In keeping with our approach of
considering leading corrections, we will truncate $\CL^{(2)}$ and $\CL^{(4)}$
to lowest order in derivatives, which means replaces tilded fields by
untilded fields, and keeping only $\CL_0^{(4)}$ in the notation of \ks.  
We stress that this procedure is not necessarily justified; 
we are considering it in an attempt 
to understand the systematics of the expansion.  

We will take a perturbative approach, consisting of inserting the lowest
order tachyon background into the equations of motion of the higher level
fields.  We will solve for the profiles of the higher level fields
numerically, and then numerically evaluate their contribution to the tension.
Expanding out the action to  the level of accuracy described above, 
we find that
$B_\mu$ and the traceless part of $B_{\mu\nu}$ decouple from the other
higher level fields, while $\beta_1$ and the trace of $B_{\mu\nu}$ are
mutually coupled. The total action will be written as 
$S= 2\pi^2 T_{25}\int \!d^{26}x\, \CL(\beta_1,B_\mu,B_{\mu\nu})$.
 We first consider the decoupled fields.

\subsec{Contribution of $B_\mu$}

We find 
\eqn\vectoract{\CL(B_\mu)= \half \partial_\mu B_\nu \partial^\mu B^\nu
+\half\left(1+{2^9 \over 3^4}\kappa \phi \right) B_\mu B^\mu 
- {4  \over 3}\kappa \phi^2 \partial_\mu B^\mu.}
Our goal is to minimize the energy of  $B_\mu$ in the presence of the 
tachyon lump background.  Asymptotically, $B_\mu$ will vanish, as this
minimizes its potential energy for $\phi= \phi_c$.  Thus we look for the
lowest energy solution to the equations of motion following from 
$\CL(B_\mu)$ which vanishes at infinity.  It is convenient at this stage
to use the approximate form $\phi = \phi_\ell$ for the tachyon background.
Only the $\rho$ component of $B_\mu$ is excited.  
Solving the equations of motion numerically yields the solution displayed
in figure 3.
\fig{$B_\rho$ profile in the presence of tachyon 
lump background.}{Bx.ps}{3truein}

The contribution of $B_\mu$ to the lump's tension is then found by 
integrating the solution
\eqn\Btens{\delta T_{24}^\ell = 
2\pi^2 T_{25}\int \!d\rho\, \CL(B_\mu) \approx  -.032 \,T_{25}.}
This represents a negligible $.6 \%$ correction to the tension.

\subsec{Contribution of traceless part of $B_{\mu\nu}$}

We first decompose $B_{\mu\nu}$ as
\eqn\Bdec{B_{\mu\nu} = \hat{B}_{\mu\nu} + {1 \over \sqrt{26}}B \eta_{\mu\nu},}
where $\hat{B}_{\mu\nu}$ is traceless.  Our definition of $B$ agrees with 
that of \ks.  Now we work out the action of $\hat{B}_{\mu\nu}$ in the
presence of the lump.  There are contributions both from the components
$\hat{B}_{\rho\rho}$ and $\hat{B}_{xx}$, where $x$ denotes coordinates
parallel to the lump.  

For $\hat{B}_{\rho\rho}$ we find
\eqn\Bpp{\CL(\hat{B}_{\rho\rho})= \half(\partial_\rho\hat{B}_{\rho\rho})^2
 + \half(1+{2^{10} \over 3^5} \kappa \phi)
(\hat{B}_{\rho\rho})^2
-{2^4 \sqrt{2}\over 3^2} \kappa [\phi \partial_\rho^2 \phi
-(\partial_\rho\phi)^2]\hat{B}_{\rho\rho}.}
The profile and tension contribution of $\hat{B}_{\rho\rho}$ are found as
above.  The profile is displayed in figure 4.

\fig{$\hat{B}_{\rho\rho}$ profile in the presence of tachyon 
lump background.}{Brhorho.ps}{3truein}
For the contribution to the tension we find
\eqn\Brhorhotens{\delta T_{24}^\ell = 
2\pi^2 T_{25}\int \!d\rho\, \CL(\hat{B}_{\rho\rho}) \approx  -.27\, T_{25},}
which corresponds to a $5 \%$ decrease in the tension.  

For $\hat{B}_{xx}$ we find for each of the 25 components
\eqn\Bxx{\CL(\hat{B}_{xx})= \half(\partial_\rho\hat{B}_{xx})^2
 + \half(1+{2^{10} \over 3^5} \kappa \phi)
(\hat{B}_{xx})^2.}
Although there is no term linear in $\hat{B}_{xx}$, this field is excited
since in deriving the Euler-Lagrange equation from $\CL(\hat{B}_{\mu\nu})$
one must respect the tracelessness of $\hat{B}_{\mu\nu}$, and this 
produces a source term for $\hat{B}_{xx}$.  
The profile is displayed in figure 5.

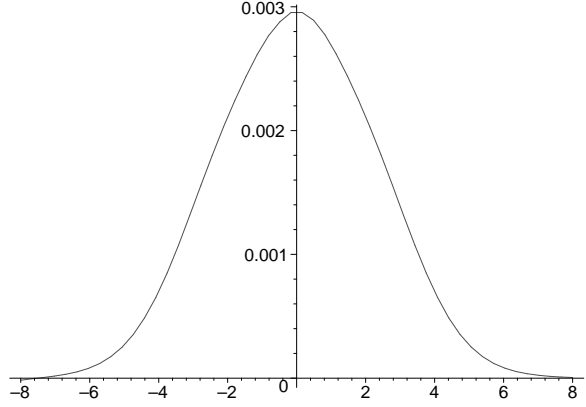
\fig{$\hat{B}_{xx}$ profile in the presence of tachyon 
lump background.}{Bxx.ps}{3truein}

The contribution of $\hat{B}_{xx}$ to the tension is 
\eqn\Bxxtens{\delta T_{24}^\ell = 
25 \cdot 2\pi^2 T_{25}
\int \!d\rho\, \CL(\hat{B}_{xx}) \approx  .01 \, T_{25},}
which corresponds to a $.2 \%$ increase.

\subsec{Contribution of $\beta_1$ and $B$}

The action for the coupled $\beta_1$, $B$ system is
\eqn\Bbeta{\eqalign{\CL(\beta_1,B)=& \half(\partial_\rho B)^2
+ \half\left[1+2({26\cdot 5^2 \over 3^5} + {2^9 \over 3^5})\kappa 
\phi\right]B^2 \cr
-&\left[{\sqrt{2} \cdot\sqrt{26} \cdot5\over 3^2}\kappa \phi^2
+ {\sqrt{2} \cdot2^4 \over \sqrt{26} \cdot3^2}\kappa
(\phi \partial_\rho^2 \phi - (\partial_\rho \phi)^2)\right]B \cr
-&\half(\partial_\rho \beta_1)^2
- \half\left[1+{2\cdot 19 \over 3^4}\kappa \phi\right]\beta_1^2
-{2 \cdot 11 \over 3^4}\kappa \phi^2 \beta_1 \cr
+&{\sqrt{2} \cdot \sqrt{26} \cdot 2 \cdot 5 \cdot 11 \over 3^5}\kappa
\phi \beta_1 B. \cr}}
Note that the auxiliary scalar $\beta_1$ has  wrong sign kinetic and
mass terms.  For this reason, the solutions for $\beta_1$ and $B$ 
in the presence of the lump will represent a saddle point of the 
energy functional \Bbeta.  This makes finding the solution difficult,
since the standard Gauss-Seidel algorithm will not converge to the 
desired solution.  In order to obtain a rough estimate of the energy
we will proceed as follows.  We insert Gaussian wave-functions 
into the equations of motion following from \Bbeta, and then vary the
height and widths so as to minimize the integrated sum of squared error
terms.  

We first need to find the boundary conditions at infinity.  These are 
given by the constant field values which  extremize the action in the 
presence of $\phi=\phi_c$.  We find
\eqn\bndycon{ B_c = .144, \quad \beta_1^c=-.127. }
Our Gaussian ansatz is then
\eqn\gauss{B = B_c+ \alpha_1 e^{-\alpha_2 \rho^2}, \quad 
\beta_1 = \beta_1^c +\alpha_3 e^{-\alpha_4 \rho^2}.}
Minimizing the error terms in the differential equations leads to the estimate
\eqn\est{\alpha_1 \approx -.2, \quad \alpha_2 \approx .15, \quad 
\alpha_3 \approx 0.}
Inserting our ansatz back into the action and integrating leads to the
following value for the contribution to the tension
\eqn\estten{\delta T_{24}^\ell =2 \pi^2 T_{25} \int \! d\rho \,
\CL(\beta_1,B) \approx T_{25}.}
This represents a $20 \%$ increase in the tension.  

Given the relatively large magnitude of this correction term, it would
clearly be desireable to improve upon the crude estimate given here.

\subsec{Corrected tension}

We can now assemble the various corrections to the tension which we have
computed.  Our zeroth order computation yielded 
\eqn\tenorig{T_{24}^\ell \approx \, 4.93 T_{25} \approx \, .78 T_{24}.}
Adding the derivative and higher level correction terms gives
(only the higher level correction from $\hat{B}_{\rho\rho}$, $\beta_1$,
$B$ are significant) 
\eqn\tencorr{T_{24}^\ell + \delta T_{24}^\ell \approx 
(4.4 -.27 + 1)T_{25}  \approx .82\, T_{24}.}
Although the precise value of our result is not significant, we note that
we have obtained a small correction in the right direction, which supports
the conjecture that the result after including higher order effects will 
converge to the expected $D24$-brane tension.

\newsec{Fluctuation Spectrum of the $D24$-brane}
In the previous sections we have seen that the level truncation scheme
in open string field theory seems to provide a reliable calculation of
$Dp$-brane tension for large $p$. It is natural to ask whether other features
of $D$-branes are also visible in this approximation. In this section we
look at the low-lying spectrum of the $D24$-brane and argue that both
the tachyon and $U(1)$ gauge field on the $D24$-brane are rather
accurately described at  the lowest non-trivial level. 

We start with the action for the tachyon and gauge field, keeping terms
up to level 2. From \ks\ this is
\eqn\action{\eqalign{S = & 2 \pi^2 T_{25} \int d^{26} x  
\biggl[ \half \p_\mu \phi \p^\mu \phi
+\half \p_\mu A_\nu \p^\mu A^\nu - \half \phi^2  \cr
& + \kappa \Bigl[  2 \phi^3 + {  2^5 \over 3^2} A_\mu A^\mu \phi 
- {2^4 \over 3^2} \bigl( 2 \p_\mu \phi A_\nu
\p^\nu A^\mu - \phi \p_\mu A_\nu \p^\nu A^\mu -
\p_\mu \p_\nu \phi A^\mu A^\nu \bigr) \Bigr]  \biggr]. \cr }}
Varying  gives
the equations of motion:
\eqn\eom{\eqalign{ - \partial^2 \phi = & \phi -
6 \kappa  \phi^2 - {2^5 \kappa  \over 3^2} A^\mu A_\mu -
{2^4 \kappa  \over 3^2} \biggl[ 3 \p_\mu A_\nu \p^\nu A^\mu +
 \p_\mu \p_\nu (A^\mu A^\nu) + 2 A_\nu \p^\mu \p^\nu A_\mu \biggr] \cr
- \p^2 A_\nu = & - {2^6 \kappa  \over 3^2} \phi A_\nu +
{2^5 \kappa  \over 3^2} \biggl[ 2 \p_\mu \phi \p_\nu A^\mu -
2 \p_\mu \p_\nu \phi A^\mu - \p_\nu \phi \p^\mu A_\mu + 
\phi \p_\mu \p_\nu A^\mu 
 \biggr]. \cr }}

To study the spectrum of small fluctuations we linearize these equations
about the lump solution $\phi_l(\rho)$ given in \applump. Taking $A_{25}=0$,
imposing $\partial^\mu A_\mu =0$, 
writing
\eqn\linfluc{\eqalign{\phi(x^\mu) = & \phi_l(\rho) + \psi_t(\rho) t(x^m) \cr
                    A_m(x^\mu) = & \psi_a(\rho) a_m(x^n) \cr}}
and substituting into \eom\ gives
\eqn\flucs{\eqalign{\psi_a \p_n \p^n a_m   = & - a_m \bigl( {\psi_a}''  -
 {2^6 \kappa \over 3^2} \phi_l  \psi_a \bigr) \cr
                    \psi_t \p_n \p^n t = & - t  \bigl( {\psi_t}'' + \psi_t -
 12 \kappa \phi_l \psi_t \bigr) \cr }}
where primes indicate derivatives with respect to $\rho$.

The masses of the fluctuations are thus given by the 
eigenvalues of the bound states of  a
one-dimensional Schrodinger equation in a potential proportional to 
$\phi_l(\rho)$: 
$m_a^2 = \lambda_a, m_t^2 = \lambda_t-1$ where $\lambda_{a,t}$ are the
eigenvalues of
\eqn\scheqn{\eqalign{- {\psi_a}'' + 
{2^6 \kappa \over 3^2}\phi_l(\rho) \psi_a = &  \lambda_a \psi_a, \cr
                     - {\psi_t}'' + 12 \kappa \phi_l(\rho) \psi_t = 
& \lambda_t \psi_t. \cr }}
A simple numerical  calculation of the minimum eigenvalues
of \scheqn\ gives
\eqn\masses{ m_t^2 \approx  -1.3, \qquad m_a^2 \approx -.06 }
in good agreement with the expected values $m_t = -1, m_a=0$. 

Note that the mechanism responsible for the $U(1)$ gauge field
is quite similar to the Randall-Sundrum mechanism for spin 2
zero modes on a domain wall \rs. This should be contrasted with the
way that the $U(1)$ gauge field zero mode arises from the
space-time point of view for D-branes and NS- IIB fivebranes in the 
superstring. 
There the zero mode arises from gauge transformations of R-R and
NS-NS tensor fields which are non-trivial in the presence of
the brane \zrefs\ whereas here the zero mode arises directly from
fluctuations of a bulk $U(1)$ gauge field. Such a zero mode would
not arise from a general action of the form  \action\ without fine
tuning. It seems likely that tachyon condensation in open string
theory naturally produces the couplings required to have an
exact zero mode. 

\newsec{Conclusions}
We have seen that $Dp$-branes can be described as lumps in the open string
field theory on a $D25$-brane, at least for sufficiently large values of $p$.
For the $D24$-brane the tension is close to the expected value and we also
find the correct spectrum of low-energy excitations on the brane, namely
a tachyon instability and an approximately massless $U(1)$ gauge field.
The $U(1)$ gauge field arises by a mechanism similar to the Randall-Sundrum
mechanism for gravity \rs.  This mechanism might have interesting applications
to brane world scenarios. 

It would be interesting to extend these results in several directions. 
$Dp$-branes for smaller $p$ should be studied to see whether one can find
reliable solutions by including higher order derivative and higher
level terms. It is also  possible that the level truncation scheme will only
be a good approximation for large $p$ and that further understanding will
require either an analytic solution or a new approximation scheme. 
In addition, 
the higher derivative and higher level terms seem to be conspiring to
freeze out the open string degrees of freedom  in the vacuum with tachyon
condensation, essentially confining open strings. It would be nice
to have a better understanding of this phenomenon (see \yi, \senpuz~ for
discussions). 

Another interesting direction is suggested by the following argument. 
Consider  a fundamental string ending on a $D24$-brane. This acts
as a source of electric flux on the $D24$-brane. When the tachyon
on the $D24$-brane condenses the flux should be confined, presumably into
a  vortex
which can be interpreted as a macroscopic string. This suggests
that after tachyon condensation 
it should be possible to directly construct macroscopic
fundamental closed strings as vortices in the open string field
theory on a $D25$-brane.

Finally, it would be
interesting to extend the analysis here to the study of both the
stable and unstable D-branes of superstring theory using
open superstring field theory on unstable $D9$-branes.

\bigskip\medskip\noindent
We would like to thank V. Balasubramanian, 
D. Kutasov and E. Martinec for  helpful 
conversations. 
This work was  supported in part by NSF Grant No.\ PHY 9901194.  

\listrefs
\end